\documentclass[11pt]{article}

\usepackage{jcappub} 
\usepackage[normalem]{ulem}
\usepackage[T1]{fontenc}
\usepackage{slashed,amsmath,amsfonts,amssymb,bbm}
\usepackage{mathrsfs,wasysym}
\usepackage{units}
\usepackage{graphicx}
\usepackage{makecell}
\usepackage{booktabs,multirow}
\usepackage{commath}
\usepackage{xspace}
\usepackage[dvipsnames]{xcolor} 
\usepackage{comment}
\usepackage{bbm}
\definecolor{darkblue}{rgb}{0,0,0.5}
\definecolor{darkgreen}{rgb}{0.0,0.5,0.2}
\definecolor{darkred}{rgb}{0.6,0,0}
\usepackage{hyperref}
\usepackage[capitalise]{cleveref}
\usepackage{orcidlink}

\renewcommand{\to}{\rightarrow}

\usepackage{tikz}
\usepackage{tikz-feynman}

\tikzfeynmanset{
yourboson/.style={
decoration={coil,aspect=0,amplitude=1.5mm,segment length=4mm,pre,pre length=0pt, post,post length=0pt},decorate}
}

\tikzfeynmanset{
yourboson2/.style={
decoration={coil,aspect=0,amplitude=1.5mm,segment length=4.1mm,pre,pre length=0pt, post,post length=0pt},decorate}
}

\subheader{\rm IFT-UAM/CSIC-26-5}

\title
{
Stochastic galactic supernova flux of semi-relativistic particles
}

\author[a,b]{D.~Alonso-González,}
\author[a]{D.~Cerde\~no,}
\author[a,c]{M.~Cerme\~no,}
\author[d]{A.D. Perez}
\affiliation[a]{Instituto de F\'isica Te\'orica IFT-UAM/CSIC, 
Cantoblanco, E-28049, Madrid, Spain}
\affiliation[b]{Departamento de F\'isica Te\'orica, Universidad Aut\'onoma de Madrid, Cantoblanco,\\ E-28049, Madrid, Spain}
\affiliation[c]{Departamento de FAIAN, ETSIN, Universidad Politécnica de Madrid, Av. de la Memoria~4, E-28040, Madrid, Spain}
\affiliation[d]{Instituto Balseiro, Centro At\'omico Bariloche, Av. Bustillo 9500, S.C. de Bariloche, 8400, R\'io Negro, Argentina}
\emailAdd{david.alonsogonzalez@uam.es}
\emailAdd{davidg.cerdeno@ift.csic.es}
\emailAdd{marina.cermeno@upm.es}
\emailAdd{andres.perez@ib.edu.ar}

\abstract
{
New exotic particles with MeV masses, such as axion-like particles or light dark matter, can be emitted from core-collapse supernovae (SNe) with semi-relativistic velocities. Due to their speed dispersion, they would arrive at Earth as an extended packet with a time spread that can be as large as tens of millennia for typical detectors. It has been argued in the literature that the superposition of packets from all galactic SNe would give rise to a smooth and stationary diffuse flux that could be observable on terrestrial experiments. In this article, we critically examine this hypothesis by carrying out a numerical simulation of the galactic history of SN explosions. We show that, although the particle packets do overlap, due to the short observational time window, each of them only contributes with a very narrow range of energies and with an intensity that depends on the SN distance. As a consequence, the energy dependence of the resulting flux is extremely sensitive to the stochastic nature of the SN population and far from smooth.
This has profound implications for the expected signature in terrestrial experiments, which displays a spectral shape that is not properly described by the smooth approximation. We develop a numerical tool to compute this stochastic galactic flux for generic semi-relativistic particles, which also allows us to explore sub-MeV particles, where the smooth diffuse flux approach does not hold. To test this framework, we revisit existing bounds on axion-like particles and fermionic dark matter, finding weaker constraints than previously reported. 
}

\begin{document}

\notoc

\maketitle

\section{Introduction}
\label{sec:intro}

In Ref.~\cite{Alonso-Gonzalez:2024ems}, we argued that axion-like particles (ALPs) with MeV masses can be produced in core-collapse supernovae (SNe) and leave the proto-neutron star (proto-NS) core with semi-relativistic velocities, resulting in a diffuse galactic flux that could be observable in neutrino water Cherenkov detectors. The same argument had been previously applied to direct detection of MeV fermionic dark matter (DM) that interacts with the Standard Model (SM) via a heavy dark gauge boson in Refs.~\cite{DeRocco:2019jti, Baracchini:2020owr} and more recent works have also exploited this idea \cite{Bhalla:2025vnq}. In this article we will collectively refer to these semi-relativistic particles as SRPs.

The existence of such a diffuse background of SRPs can be explained as follows. After being produced in the first seconds of core-collapse SN, MeV particles would escape the proto-NS core with semi-relativistic velocities. SRPs would travel at different speeds, depending on the energy at which they are produced, leading to a packet that is very spread in time when it arrives at Earth (approximately $10^2$ to $10^4$~years for the typical energy ranges that are observable in DM direct detection and neutrino experiments). With a rate of approximately 1.6 galactic SNe per century~\cite{Rozwadowska:2020nab}, it is expected that the packets from $\sim 10-100$ SNe overlap. The resulting diffuse flux of SRPs is then computed by averaging over the distance at which the SNe exploded and the position-dependent probability density of SNe in the Milky Way, and it is therefore smooth and stationary.

However, there are two important aspects that have been overlooked in the argument above and that make previous computations of the diffuse flux inaccurate. First, the difference in arrival time of SRPs produced in a single SN is a function of the energy at which they are produced. Due to the very small observation time window (of the order of $1-10$~yr) compared with the time scale of the whole SRP packet, which can reach values of $\sim 10^3$~yr, we only observe a very small slice of each individual SRP spectrum (the energy and width depend on the distance of that specific SN and the SRP mass). Thus, although the SRP packets do overlap at Earth, the resulting flux is not smooth. Second, for some combinations of SN distance, explosion time, and SRP mass, the fraction of the SRP packet with energies in the observation range has already crossed the Earth.\footnote{It should be noted that the analysis is limited to galactic sources since the contribution from extra galactic SNe is expected to be two orders of magnitude smaller. This is analogous to the comparison between the average of the neutrino detection rate
from Milky Way supernovae and that from the diffuse supernovae neutrino background \cite{Beacom:2010kk}. This sets a limit to the farthest SN and to the travel time of SRPs.} This leads to a reduction of the total flux, which is more pronounced for light SRPs.

In order to investigate these effects, in this article we carry out a numerical simulation of the core-collapse SN history in our galaxy, computing the resulting SRP flux at Earth after taking into account the position and time of explosion of each individual event. We demonstrate that, rather than a smooth flux, SRPs give rise to a {\em stochastic} flux that is strongly dependent on the history of galactic SNe. This numerical approach evidences that the expected energy spectrum in Earth detectors can differ significantly from the naive treatment. It also allows us to explore SRPs lighter than 1~MeV, where the usual smooth flux approach is not valid. We study the implications that this has on the detection of these particles, concentrating on the particular cases of MeV ALPs and fermionic DM particles, which had already been discussed in the literature within the smooth diffuse flux approximation.  We show that the latter led to overly restrictive bounds.

This article is organised as follows. In \cref{sec:sprflux}, we analyse the flux of SRPs with MeV-scale masses arriving at Earth from past galactic SNe. We first review the standard smooth approximation (\cref{subsec:diff}), and then present a more realistic treatment based on a numerical simulation of SN events in our galaxy (\cref{subsec:simulation}). In \cref{sec:alps}, we study the particular case of ALPs and their detectability in neutrino water Cherenkov experiments. In \cref{sec:darkphotons}, we consider the case of fermionic DM interacting via dark photons and study the prospects for its detection in direct detection experiments. Finally, in \cref{sec:conclusions} we present our conclusions.

\section{Diffuse flux of SRPs at Earth from past galactic SNe}

To characterize the SN environment, we adopt a spherically symmetric simulation of an $18$ solar mass, $M_\odot$, progenitor performed with the \textsc{AGILE-BOLTZTRAN} code~\cite{agile1,agile2}. Specifically, we use the profiles of temperature, density, and electron and muon fractions reported in Ref.~\cite{Fischer:2021jfm}, one second after core bounce, when the highest temperature is reached and particle production is maximised. The spectrum of the SRPs is then obtained by integrating over a 1.5 second interval, from $t_{\rm{min}} = 0.5\,\rm{s}$ to $t_{\rm{max}} = 2\,\rm{s}$, since within this period the profiles remain approximately unchanged (see, e.g., Fig.~7 in Ref.~\cite{fischer2012}). This approach results in a good, yet conservative, estimation of the total SRP flux as well as the signal induced in the detectors accordingly. For simplicity, we assume that our SRPs do not modify the SN profiles, which has been shown to hold for massless axions in Ref.~\cite{Fischer:2021jfm}.

In this work, we use 1D spherically symmetric SN profiles.  3D simulations, however, indicate that the dynamics of the explosion can be anisotropic, leading to directional variations of the neutrino signal as large as 20\% ~\cite{Tamborra:2014aua,Vartanyan:2019ssu}. As an approximation, one can assume the same uncertainty in the flux of exotic particles. However, it is important to highlight that a detailed assessment of the anisotropy impact on the emission of new particles can be model dependent, therefore, dedicated simulations including their particular interactions with the environment would be needed.

Another source of uncertainty is the SN initial mass function (IMF). For simplicity and to compare with previous results, we model all galactic SNe using the profiles of an $18\,M_\odot$ progenitor. In reality, most explosions are expected to originate from stars of lower mass~\cite{Sukhbold:2015wba}. The peak temperatures and densities of these lighter progenitors one second after bounce are typically smaller than in the $18\,M_\odot$ case. For example, simulations of an $8\,M_\odot$ progenitor predict values about 20\% below those employed in our reference model~\cite{Calore:2021hhn}. Conversely, approximately 15\% of core-collapse events correspond to progenitors more massive than $18\,M_\odot$, which can reach substantially higher temperatures and densities. In the context of the detection of ultralight ALPs from extragalactic origin, in Ref.~\cite{Calore:2021hhn} the uncertainty on the bounds due to the SN IMF was found to be subleading and estimated at $\sim 7\%$. A realistic estimate of the variation in the resulting SRP flux would require a systematic study including the full progenitor mass distribution and corresponding simulations, which we do not explore in this work.

\label{sec:sprflux}

\subsection{Smooth approximation}
\label{subsec:diff}

It has been previously argued that, once they leave the SN, massive SRPs could form a diffuse galactic flux \cite{Alonso-Gonzalez:2024ems,DeRocco:2019jti,Baracchini:2020owr,Bhalla:2025vnq}. Since they are produced with a range of energies, SRPs will propagate with a spread in velocities, resulting in a difference in arrival time at Earth between the high and low-velocity SRPs. A SRP with mass $m$ produced with energy $E$ in a SN at a distance $d$ would take a time $t_{\rm SRP}(E)$ to reach the Earth, with
\begin{equation}
 t_{\rm SRP}(E)=\frac{d}{v(E)}=\frac{d}{\sqrt{1-\left({m}/{E}\right)^2}}\   .
 \label{eq:timedelay}
\end{equation}
The duration of the observable part of a given SRP packet, $\Delta t_{\rm SRP}$, is determined by the experimental energy window, $[E^{\rm min}, E^{\rm max}]$, as $\Delta t_{\rm SRP}= t_{\rm SRP}(E^{\rm min}) - t_{\rm SRP}(E^{\rm max})$.
For example, for a SN located at the Galactic Centre (GC), $d=8.2$~kpc \cite{Leung_2022}, a SRP of mass $m=1-100$~MeV, and an energy region of interest of $30-100$~MeV (typical of neutrino Cherenkov detectors), one obtains $\Delta t_{\rm SRP}\sim 10^2-10^4$~yr.
Since we expect a galactic SN rate of $1.63 \pm 0.46$ events per century~\cite{Rozwadowska:2020nab}, it was argued that this would result in a near-constant galactic flux at any time at Earth due to the overlap of several SNe. 
In previous works it was pointed out that for SRP masses smaller than 1~MeV, the time spread $\Delta t_{\mathrm{SRP}}$ becomes shorter, and the above approximation progressively breaks down since fewer SNe contribute simultaneously to the flux (notice that SRPs with those masses are highly relativistic).

The galactic SN rate profile is usually modelled following the stellar distribution in our galaxy~\cite{Adams:2013ana}
\begin{equation}
    \frac{dn_{\rm SN}}{dt} = A \, \mathrm{e}^{-{r}/{R_d}} \, \mathrm{e}^{-{|z|}/{H}}\, .
    \label{eq:SNrate}
\end{equation}
For Type II SNe, $R_d=2.6$~kpc, $H=300$ pc~\cite{McMillan:2016jtx} and the normalization factor is $A=6.40\times10^{-4}$ kpc$^{-3}$ yr$^{-1}$.

The total flux of SRPs at Earth is obtained by integrating the galactic SN rate and the SRP spectrum produced by a single SN, ${dN}/{dE^{\rm Earth}}$, over the whole galaxy to obtain
\begin{align}
    \frac{d\Phi}{dE^{\rm{Earth}}}&=\int_{-\infty}^{\infty} \int_0^{2\pi} \int_0^{\infty} \frac{dn_{\rm SN}}{dt} \, \frac{r}{4\pi\, (\vec{r} - \vec{R}_E)^2} \frac{dN}{dE^{\mathrm{Earth}}}\, dr \, d\theta \,  dz  \nonumber\\
    &= \bar{\Phi} \, \frac{dN}{dE^{\mathrm{Earth}}}\, ,
    \label{eq:diff_factor}
\end{align}
where $|\vec{R_E}|=8.2$~kpc is the Earth distance with respect to the GC, $z_E=20.8$~pc is its position above the mid-plane of the disk, and we assume that ${dN}/{dE^{\rm Earth}}$ is isotropic. The diffuse flux factor $\bar{\Phi} = 1.3 \times 10^{-55} \; \mathrm{cm^{-2}  \; s^{-1}}$, with a $\sim 30 \%$ uncertainty given by current errors in the SN rate and distribution, depends only on the SN profile rate and not on the details of the SRP model. To compute the total flux within an observation interval $\Delta t_{\rm obs}$, we can define a diffuse fluence factor as $\bar{\mathcal{F}} = \Delta t_{\rm obs} \, \bar{\Phi}$, since $\bar{\Phi}$ is constant.

However, two key effects are missing from the previous argument. First, since the observational time window in the experiment, $O(1-10)$~yr, is very small compared with the overall duration of the SRP packet (given by the time spread $\Delta t_{\rm SRP}\gtrsim10^2$~yr), only a very narrow slice of the spectrum of each SN would be observed at the detector. The range of energies and spectral width are a function of both the progenitor location and the SRP mass. Consequently, even though SRP packets from different galactic SNe overlap at the same time at Earth, only a small portion of the spectrum of each individual SN would be detected. Therefore, the effective averaging over all the spectra done in the smooth diffuse flux approximation does not hold.

Second, for certain combinations of the progenitor distance, explosion time and SRP mass, it is possible that the relevant part of the SRP packet has already crossed the Earth before the observational time window. In such cases, the remaining SRPs have too little energy to produce signals above the detector threshold. This leads to a reduction of the total expected detectable flux.

\subsection{Stochastic treatment and numerical simulation}
\label{subsec:simulation}

To determine the exact SRP flux at Earth produced in the galaxy, one would need the location and explosion time of all past SN. Since the SN galactic history is not known, we perform a set of numerical simulations by sampling the times and locations of SNe, and compute the aggregated SRP flux. The code is publicly available on GitHub\footnote{\href{https://github.com/AndresDanielPerez/SGaSNoF}{https://github.com/AndresDanielPerez/SGaSNoF}} as a Python implementation.

We consider a time window of $9\times10^5$~yr, which corresponds to the time that a particle created at $d=|\vec{R_E}|+30$ kpc with mass $m$ and energy $E=1.01\,m$ needs to reach Earth. Given the above-mentioned rate of galactic SNe, we expect an average of $\lambda_{\rm SN}=14670$ SN events. Then, the number of samples per simulation is drawn from a Poisson distribution with parameter $\lambda_{\rm SN}$, that are located in time according to a uniform distribution in the considered time window.

\begin{figure}[!t]
  \centering
  \includegraphics[width=0.8\textwidth]{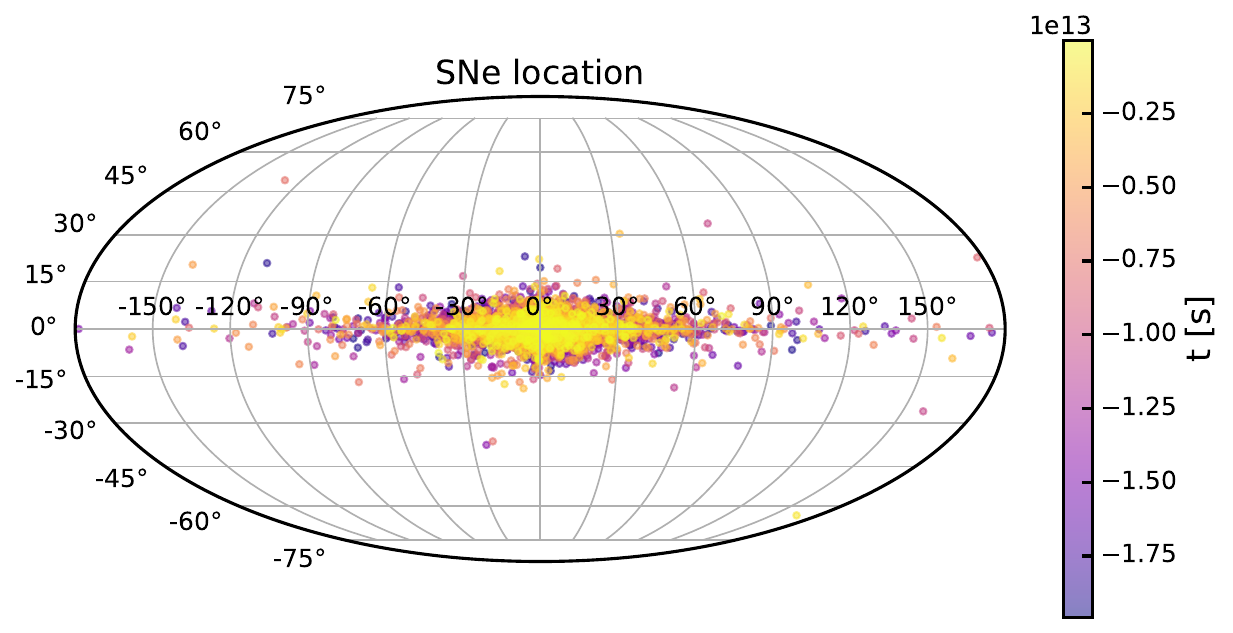}
  \caption{Example of the location in galactic coordinates of the last 10000 SNe in a simulation. The colour bar shows the time of each event.
  }
  \label{fig:SNlocation}
\end{figure}

The spatial location of each event is determined following the galactic SN rate profile~\cite{Adams:2013ana} shown in \cref{eq:SNrate}. Specifically, in cylindrical coordinates $(r, \theta, z)$, with the GC at the origin, $r$ is drawn from an exponential distribution with scale $R_d$, $\theta$ from a uniform distribution within the interval $[0, 2\pi)$, and $z$ from a Laplace distribution with scale $H$. In \cref{fig:SNlocation} we show an example with the location of the latest 10000 SNe in galactic coordinates.

Once we have the location and time of the explosion of the SN galactic history, the SRP flux at Earth is obtained by summing over each contribution,
\begin{equation}
    \frac{d\Phi}{dE^{\rm{Earth}}}(t)=\sum_k \, \frac{1}{4\pi\, d_{k}^2} \,\frac{dN_k}{dE^{\mathrm{Earth}}}\, \delta(t-t_{\rm arr}(E^{\mathrm{Earth}})) \,,
\label{eq:stoch_factor}    
\end{equation}
where $k$ denotes the $k$-th SN in the simulation, $d_{k}$, $t_{k}$, and ${dN_k}/{dE^{\mathrm{Earth}}}$ are the distance to Earth, the time, and the total SRP spectrum of the $k$-th SN, respectively. The delta function, written in terms of the time of arrival, $t_{\rm arr}(E^{\mathrm{Earth}}) = t_{k}+t_{\rm SRP}(E^{\mathrm{Earth}})$, takes into account that SRPs with different energies reach Earth at different times (see \cref{eq:timedelay}), therefore only a part of the spectrum contributes at any moment. Then, to compute the total flux, we just need to integrate within an observation interval $[t_0, t_0+\Delta t_{\rm obs}]$ 
\begin{align}
    \int_{t_0}^{t_0+\Delta t_{\rm obs}} \frac{d\Phi}{dE^{\rm{Earth}}}(t) \, dt \, &= \, \sum_k \, \frac{1}{4\pi\, d_{k}^2} \,\frac{dN_k}{dE^{\mathrm{Earth}}}\, \Theta(E^{\mathrm{Earth}} - E_k^{\rm{min}}) \, \Theta(E_k^{\rm{max}} - E^{\mathrm{Earth}}) \notag  \\
     &= 
     \frac{dN}{dE^{\mathrm{Earth}}} \sum_k \mathcal{F}_{k}(E^{\mathrm{Earth}}) \notag  \\
     &= \mathcal{F}(E^{\mathrm{Earth}}) \, \frac{dN}{dE^{\mathrm{Earth}}}\, ,
\label{eq:stoch_factor2}    
\end{align}
where the Heaviside step functions select the energy range $[E_k^{\rm{min}}, E_k^{\rm{max}}]$ that reaches Earth in the observation interval. Taking $t_0 = 0$, the minimum and maximum energies are calculated as $E_k^{\rm{min}} = E_{\rm SRP}(t_{k}+\Delta t_{\rm obs},d_{k}, m)$ and $E_k^{\rm{max}} = E_{\rm SRP}(t_{k},d_{k}, m)$ using \cref{eq:timedelay}.

Finally, in the last steps of \cref{eq:stoch_factor2}, we have assumed that all SNe produce the same spectra, ${dN}/{dE^{\mathrm{Earth}}}$, and therefore we can factorize it. This is justified
because we adopt identical temperature and density profiles for all explosions, corresponding to
one-dimensional simulations of an $18\,M_\odot$ progenitor, as described in \cref{sec:sprflux}. Under these considerations, $\mathcal{F}_{k}(E^{\mathrm{Earth}})$ is the fluence factor produced by each SN. Then, the total stochastic fluence factor $\mathcal{F}(E^{\mathrm{Earth}})= \sum_k \mathcal{F}_{k}(E^{\mathrm{Earth}})$ depends not only on the galactic SN history, but also on the SRP mass. Additionally, notice that, unlike the smooth diffuse flux factor in \cref{eq:diff_factor}, the stochastic fluence is not constant in $E^{\mathrm{Earth}}$, since each SN contributes in a particular energy window $[E_k^{\rm{min}}, E_k^{\rm{max}}]$.

\begin{figure}[!t]
  \centering
  \includegraphics[width=0.51\textwidth]{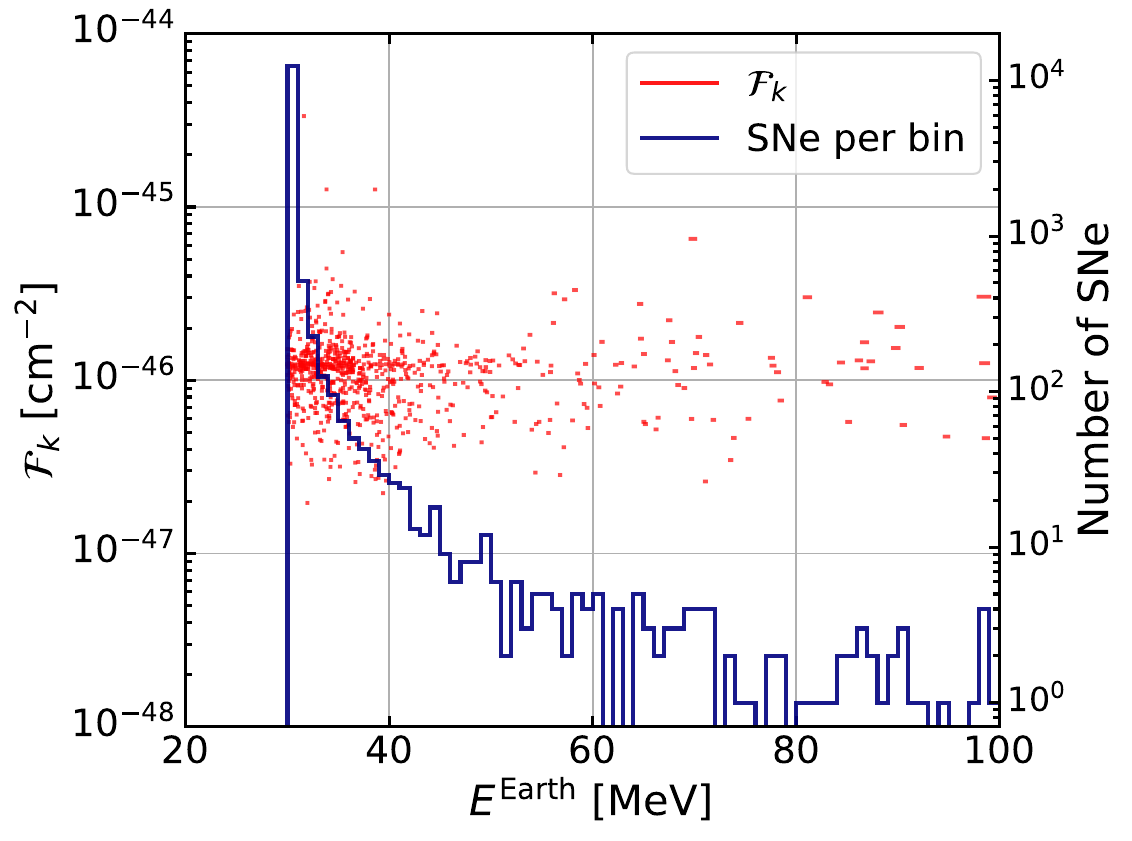}
  \includegraphics[width=0.48\textwidth]{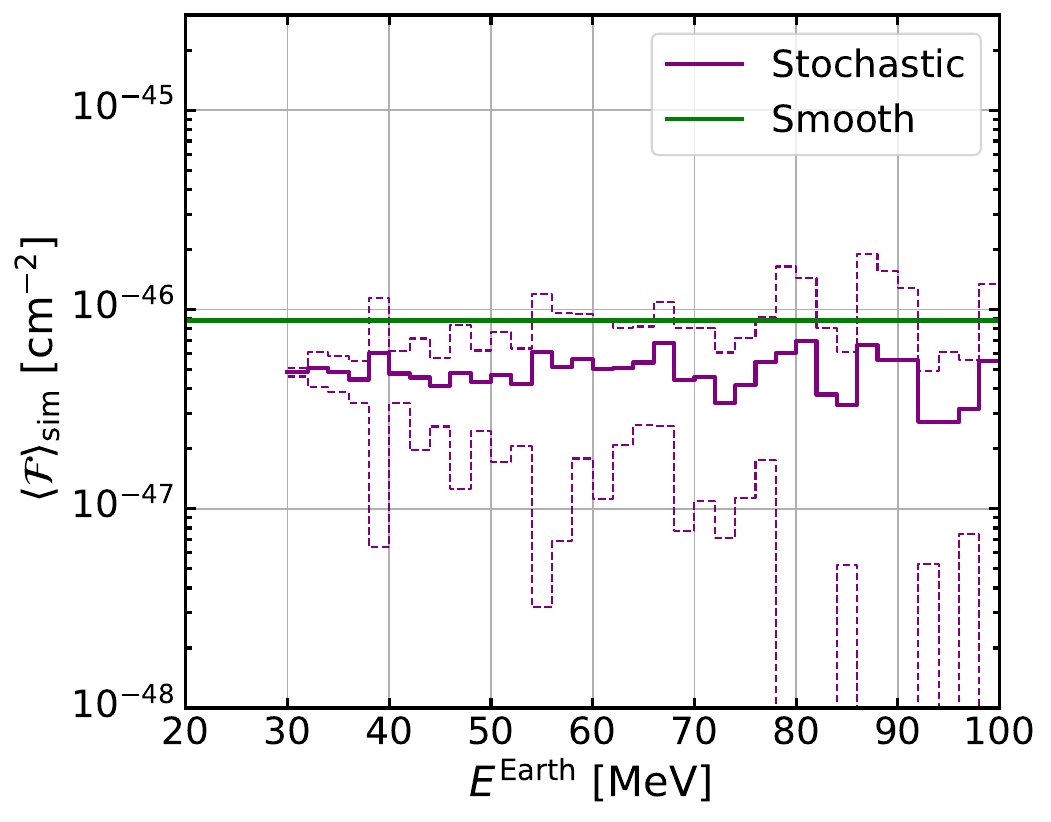}
  \caption{Left: fluence factor $\mathcal{F}_{k}$ for each SN as a function of the energy at Earth in a simulation example (red lines). The weight value and the energy range of each event depend on the SRP mass, as well as on both the location and the time of the SN. The blue histogram represents the number of SNe that contribute to each energy range. Right: smooth
  $\bar{\mathcal{F}}$ in green, and the stochastic fluence factor integrated in 2 MeV energy bins and averaged over the 20 SN galactic histories $\left\langle \mathcal{F} \right\rangle_{\rm sim}$ as defined in \cref{eq:medE} in purple. For the latter, the line corresponds to the mean value while the dashed lines are the 1$\sigma$ band, taking 20 simulated galactic SN histories. In both panels we consider a SRP of mass $m=30$~MeV and an observation time window of $\Delta t_{\rm obs}=20$~yr.} 
  \label{fig:SNcontribution}
\end{figure}

In the left panel of \cref{fig:SNcontribution}, we show as red curves the fluence $\mathcal{F}_{k}$ for all SNe in one simulation, considering an SRP with a mass of 30~MeV and an observation time window of $\Delta t_{\rm obs}=20$ yr. We can clearly see that each SN contributes to a very narrow energy range, which means that only a slice of each individual SRP spectrum could be measured by a reasonable experiment. As explained before, this follows from the fact that $\Delta t_{\rm obs} \ll \Delta t_{\rm SRP}$. However, a large number of SN contribute to the total fluence, as illustrated by the blue histogram. For SRP energies with values closer to their mass, the number of SN increases significantly, while the energy range of each $\mathcal{F}_{k}$ decreases. That region corresponds to the contribution of older SNe, where only a small flux of particles with low velocities reaches Earth today.

We compute the total stochastic fluence factor $\mathcal{F}$ as a histogram with 2 MeV bins. Since each SN galactic history results in a different total fluence, we simulated 20 SN galactic histories and computed $\mathcal{F}$ for each of them, and averaged over the all galactic histories, normalized by the bin width,
\begin{equation}
    \left\langle \mathcal{F}(E_j) \right\rangle_{\rm sim} \; = \; \frac{1}{\Delta E_j} \, \left\langle \int_{E_j}^{E_{j+1}} \mathcal{F}(E)\, \mathrm{d}E \right\rangle_{\rm sim} \, ,
    \label{eq:medE}
\end{equation}
where the index $j$ labels the energy bins. In the right panel of \cref{fig:SNcontribution} this is represented as a solid purple line, while the dashed purple lines indicate the standard deviation among different realizations, reflecting the stochastic variability of the galactic SN population. For comparison, the green solid line corresponds to the smooth fluence factor, $\bar{\mathcal{F}}$.

From both panels of \cref{fig:SNcontribution}, we can see the differences between the simulation approach and the smooth diffuse flux approximation. While for the latter one assumes that a few SN contributes with the entire spectrum, we can see from the left panel that the opposite situation holds on the numerical simulations: the entire SN history has to be taken into account since each one contributes with a small slice of its SRP spectrum For most SNe, the majority of the SRP packet has already crossed the Earth before the observational time window, leading to a reduction of the total expected detectable flux. This is illustrated by the fact that $\left\langle \mathcal{F}(E_j) \right\rangle_{\rm sim}$ is a factor $\sim 2$ smaller than $\bar{\mathcal{F}}$, as can be seen in the right panel.

Finally, it is important to note that the stochastic nature of the simulations produces significant fluence variations for different iterations. These variations are especially relevant for $E \gg m$ where fewer and more recent SN contribute to the flux and therefore their particular location and time impact the total fluence more significantly. This effect is more pronounced for low-mass particles, where the reduced temporal overlap among the different SRP packets implies that the assumptions of the smooth flux approximation are not satisfied, making a stochastic treatment necessary.

\section{ALPs from SN}
\label{sec:alps}

In Ref.~\cite{Alonso-Gonzalez:2024ems} we derived constraints on the ALP-proton coupling from the non-observation of a diffuse galactic SN ALP flux in Super-Kamiokande (SK), using the smooth flux approach. In this section, we re-evaluate these limits considering the numerical simulation of SN events derived before.

If ALPs couple to nucleons and other hadrons through the interaction Lagrangian~\cite{PhysRevD.40.652,DiLuzio:2020wdo,Chang:1993gm, Lella:2024dmx}
\begin{align}
    \mathcal{L}_{int} =\ & g_a \, \frac{\partial_{\mu} a}{2 \, m_N} \, \bigg[ 
     C_{ap} \, \bar{p} \, \gamma^{\mu} \, \gamma_5 \, p 
    + C_{an} \, \bar{n} \, \gamma^{\mu} \, \gamma_5 \, n \notag  + \frac{C_{a \pi N}}{f_{\pi}} \, \left( i \, \pi^+ \, \bar{p} \, \gamma^{\mu} \, n - i \, \pi^- \, \bar{n} \, \gamma^{\mu} \, p \right) \notag \\
    & + C_{a N \Delta} \, \left( \bar{p} \, \Delta^+_{\mu} + \bar{\Delta^+_{\mu}} \, p 
    + \bar{n} \, \Delta^0_{\mu} + \bar{\Delta^0_{\mu}} \, n \right) 
    \bigg],
    \label{eq:Laxion}
\end{align}
they can be produced in the proto-NS core via nucleon-nucleon Bremmstrahlung, $N N \to N N a$, and pion-ALP conversions, $\pi\, N \rightarrow N\, a$, see Refs.~\cite{RAFFELT19901, Raffelt:1993ix, Raffelt:1996wa, Giannotti:2005tn, Fischer:2016cyd, Carenza:2019pxu,PhysRevLett.60.1793,PhysRevLett.60.1797,burrows1989,PhysRevD.42.3297,Carenza:2019pxu,Carenza:2020cis,Fischer:2021jfm,Choi:2021ign, Lella:2022uwi, Lucente:2022vuo, Lella:2023bfb, Carenza:2023lci, Chakraborty:2024tyx, Lella:2024dmx, Alonso-Gonzalez:2024ems} for more details. In Eq.~\eqref{eq:Laxion}, $p$ denote the proton, $n$ the neutron, $\pi$ the pion, and $\Delta$ the $\Delta$-baryon. The coupling $g_a$ is a dimensionless constant which is related to the high energy ALP scale $f_a$ as $g_a =~m_N / f_a$, where $m_N =938$~MeV is the nucleon mass. The pion decay constant is $f_{\pi}=~92.4$~MeV and $C_{aN}$ are model-dependent $\mathcal{O}(1)$ ALP-nucleon couplings with $N=~p, n$. The ALP-pion-nucleon and the ALP-nucleon-$\Delta$ baryon couplings can be written as $C_{a \pi N} = (C_{ap} - C_{an}) / \sqrt{2} g_A$~\cite{Choi:2021ign}, and $C_{a N \Delta} = - \sqrt{3}/2 (C_{ap} - C_{an})$, where $g_A \simeq 1.28$~\cite{PDG:2022} is the axial coupling. We define the ALP-proton and ALP-neutron coupling as $g_{aN} = g_a \, C_{aN}$, for convenience.

For couplings above $g_{aN} \sim 10^{-8}$, ALPs are diffusively trapped inside the proto-NS core, and absorption effects given by $N\, N\, a \rightarrow N\, N$ and $N\, a \rightarrow N\, \pi$ processes become significant, as it has been shown in Refs.~\cite{Lella:2023bfb, Alonso-Gonzalez:2024ems}. In this context, the spectral fluence of ALPs at a detector located far from the SN resulting from an isotropic production spectrum is given by
\begin{equation}
    \frac{dN_a}{dE_a^{* \infty}} = \int_{t_{\rm min}}^{t_{\rm max}} dt \int_0^{\infty} dr\, \alpha(r)^{-1} 4 \pi r^2 \, \left\langle e^{-\tau\left(E_a^*, t, r\right)} \right\rangle \frac{d^2 n_a}{dE_a dt} \left( r, t, \alpha(r)^{-1} E_a^{* \infty} \right) \, ,
\label{eq:fluence}    
\end{equation}
where $E_a^{* \infty}=\alpha(r) E_a$ is the observed energy at infinity, redshifted relative to the local energy $E_a$, through the lapse function $\alpha(r) \leq 1$, which depends on the radial position with respect to the centre of the proto-NS core, $r$, and encodes the effects of the gravitational potential (see Ref.~\cite{Alonso-Gonzalez:2024ems} for more details).

The total ALP production rate in the SN, ${d^2 n_a}/{dE_a dt}$, accounts for the previously mentioned processes, whose production rates can be found in Refs.~\cite{RAFFELT19901, Raffelt:1993ix, Raffelt:1996wa, Giannotti:2005tn, Carenza:2019pxu,Carenza:2020cis, Choi:2021ign} (Ref.~\cite{Carenza:2023wsm} provides a very useful compilation of these expressions). The exponential suppression term, $\left\langle e^{-\tau\left(E_a^*, t, r\right)} \right\rangle$, encodes the absorption effects, relevant in the trapping regime. Here, $\tau\left(E_a^*, t, r\right)$ is the optical depth evaluated at $E_a^* = E_a^{\rm loc} \alpha(r)/\alpha(\sqrt{r^2+s^2+2rs\mu})$,
which is the ALP energy after accounting for the gravitational redshift between the ALP production and absorption points. Details on the computation of this quantity can be found in Refs.~\cite{Carenza:2023wsm, Carenza:2023lci}. Note that both radial and time dependences in ${d^2 n_a}/{dE_a dt}$, $\tau$, and $\alpha$, arise from their dependences on the proto-NS temperature, density, and other related quantities.

\begin{figure}[!t]
  \centering
  \includegraphics[width=0.51\textwidth]{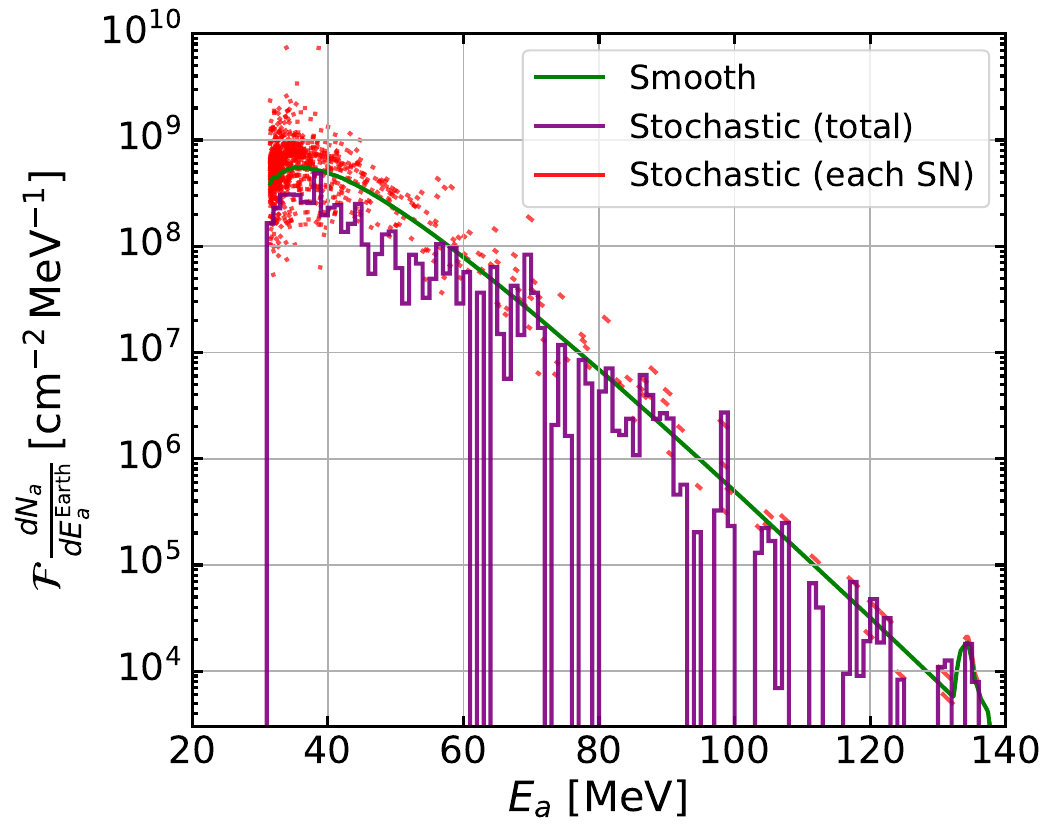}
  \includegraphics[width=0.48\textwidth]{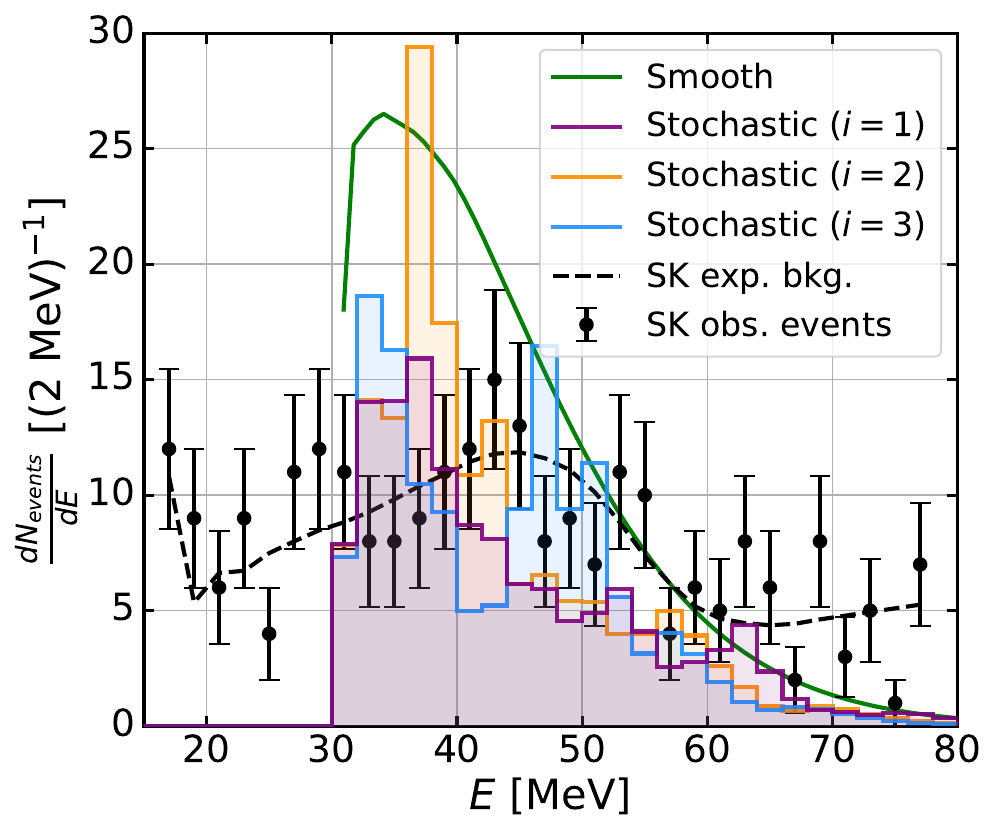}
  \caption{Left: total fluence of ALPs at Earth computed with either the smooth (green line) or the stochastic (purple and red lines) approach for an observation time window of $\Delta t_{\rm obs}=20$~yr and a single SN history. Right: expected event rate in SK phase IV via $a\ p\rightarrow p\ \gamma$, considering the smooth diffuse approximation (green line) and the stochastic approach (purple, orange and blue lines) for 3 different galactic SN history simulations. Expected events from SM processes (black dashed line) and observed events (black points with error bars) are taken from  Ref.~\cite{Super-Kamiokande:2021jaq}. Both panels consider $m_a=30$~MeV and $g_{ap}=9.4\times 10^{-5}$.
  }
  \label{fig:ALPflux}
\end{figure}

Using the ALP spectra from a single SN computed according to \cref{eq:fluence}, we can determine the total ALP fluence at Earth from past galactic SNe by inserting it into \cref{eq:diff_factor} and \cref{eq:stoch_factor2} for the smooth and stochastic approaches, respectively. The left panel of \cref{fig:ALPflux} shows an example for $m_a = 30$~MeV, $g_{a\gamma} = 9.4 \times 10^{-5}$, and an observation time window of $\Delta t_{\rm obs} = 20$~yr, similar to the exposure of the first four SK phases. To compute the total stochastic fluence, we integrated all individual SN contributions in 1~MeV bins. Since we only show the results for a single galactic SN history simulation, some bins are not populated while others display a larger fluence than the smooth case. Importantly, however, both approaches follow the same overall trend.

ALPs can be detected through their interactions with free protons in neutrino water Cherenkov detectors, producing photons via the process $a + p \rightarrow p + \gamma$, as first proposed in Refs.~\cite{Alonso-Gonzalez:2024ems,Alonso-Gonzalez:2024spi}. The resulting differential photon spectrum at an experiment with $N_t$ targets is then given by

\begin{equation}
    \frac{d\, N_\gamma}{dE_\gamma}=N_t \int_{E_a^{\rm min}(E_\gamma)}^{E_a^{\rm max}(E_\gamma)} dE_a^{\mathrm{Earth}}\frac{d\Phi_a}{dE_a^{\mathrm{Earth}}} \frac{d\, \sigma_{ap\rightarrow p\gamma}}{d\, E_\gamma},
\label{eq:photon_spectrum}
\end{equation}
where ${d\, \sigma_{ap\rightarrow p\gamma}}/{d\, E_\gamma}$ is the photo-production cross section. Further details on the computation of the cross section can be found in Ref.~\cite{Alonso-Gonzalez:2024ems}.

In the right panel of \cref{fig:ALPflux}, we show the expected event rate at SK phase IV (exposure 22.5 × 2970 kton days) considering $m_a=30$ MeV and $g_{ap}=9.4\times 10^{-5}$. From the $a\ p\rightarrow p\ \gamma$ interaction we expect $E_{\gamma} \sim E_{a}$ for the relevant energy range, and therefore the photon signal peaks at energies $\sim35$ MeV. \footnote{
For ALPs with masses of order 1 MeV or below (effectively massless in this context), the peak of the energy spectrum lies around $20-30$~MeV, since these particles are produced with energies much larger than their masses, as can be seen in Ref.~\cite{Alonso-Gonzalez:2024spi}.}  
In this case, we present the results for $i=3$ different galactic SN history simulations. 
The expected background and observed events (in black) have been taken from Ref.~\cite{Super-Kamiokande:2021jaq}. The reconstructed energy region $E_{\rm rec}=[16,80]$ MeV was optimized to search for the DSNB by identifying positrons from inverse beta decay interactions. Since the Cherenkov signal generated by one ultra-relativistic positron is indistinguishable from the one produced by a photon, we apply the same dataset to search for photons produced by ALP interactions, focusing on the signal region without neutron coincidence.

For the smooth approximation, this benchmark clearly exceeds the SK observation and would be excluded. However, the situation for the stochastic signal is not obvious: although it is $\sim 2$ times smaller on average, individual stochastic realizations can differ significantly in spectral shape, as illustrated by the three examples shown in the right panel of \cref{fig:ALPflux}. The signal shows peaks in different energy bins, whose position and amplitude depend on the specific galactic SN history. In particular, some of those peaks can even exceed the smooth expectation. This highlights that while the smooth diffuse approximation describes only an average behaviour, the stochastic treatment captures the discrete nature of galactic SN events, which is crucial for the interpretation of the data.

\begin{figure}[!t]
  \centering
  \includegraphics[width=0.49\textwidth]{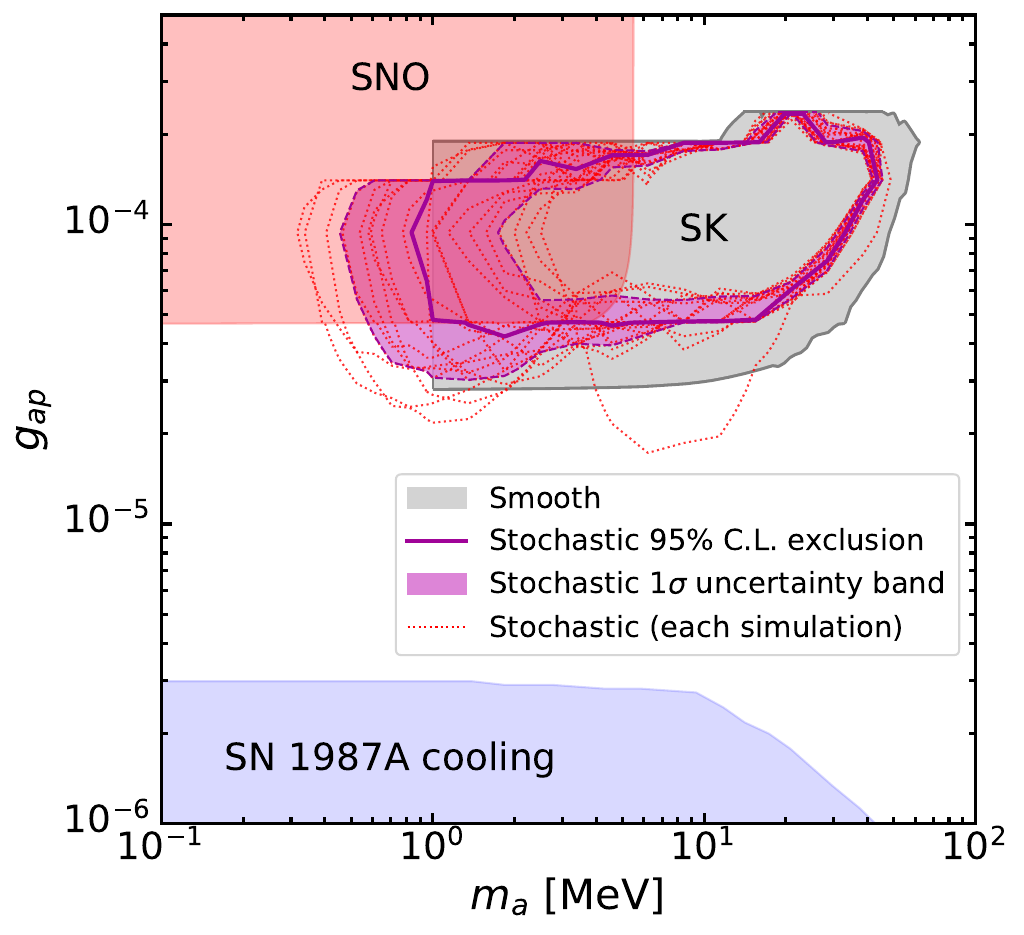}
  \includegraphics[width=0.49\textwidth]{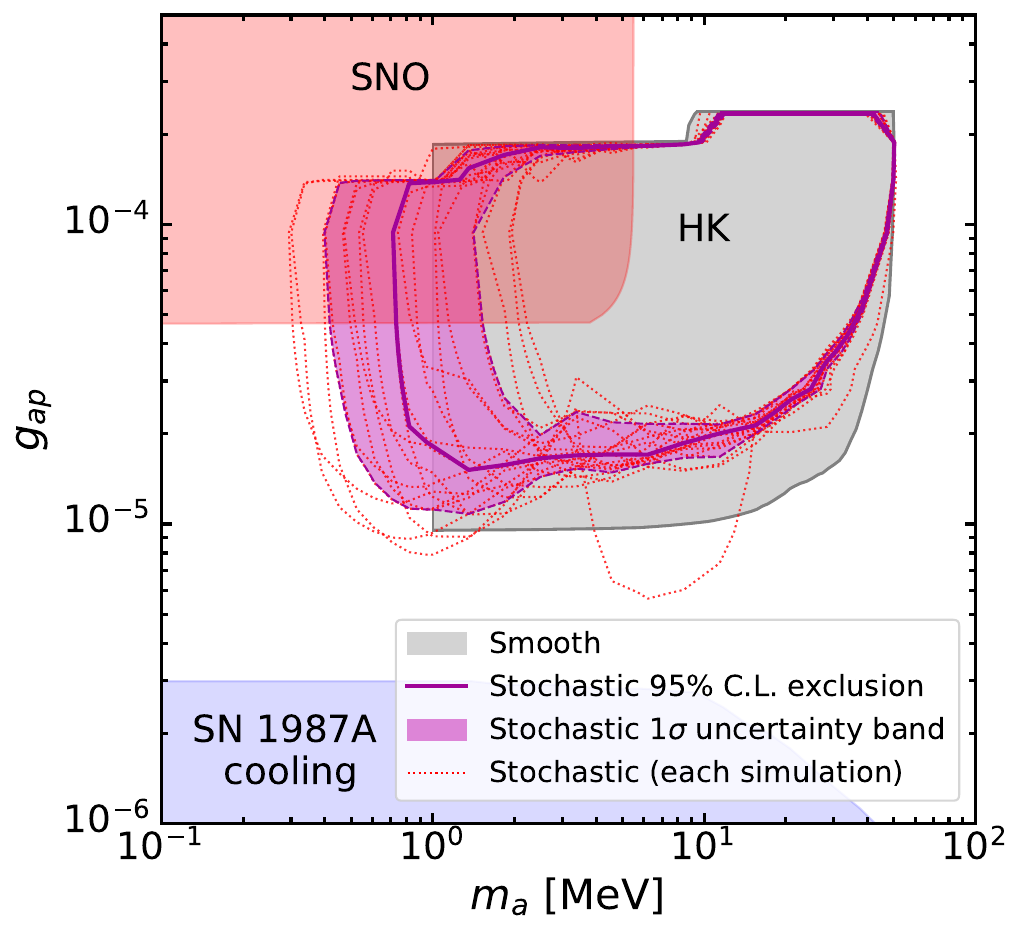}
  \caption{Current (left) and projected (right) bounds on the ALP parameter space, using the smooth diffuse approximation (gray), and the stochastic approach (purple), including a 1$\sigma$ uncertainty band from 20 SN galactic history simulations. Complementary bounds are also shown from expected solar axion events in SNO~\cite{Bhusal:2020bvx} (red) and from SN 1987A cooling~\cite{Lella:2023bfb} (light blue).
  }
  \label{fig:ALPlimits}
\end{figure}

In order to set constraints in the ALP ($g_{ap}$, $m_a$) parameter space, we follow the same statistical procedure detailed in Ref.~\cite{Alonso-Gonzalez:2024ems} over 20 SN galactic history simulations (the event rate of each one is shown in Appendix~\ref{sec:extraplots}). We compared the observed number of events with the expected signal and background yields through a profiled log-likelihood ratio test, considering a binned analysis in the reconstructed energy signal range mentioned before, with bin width of 2 MeV. Assuming that the test statistic follows a $\chi^2$ distribution, we obtained a $\chi^2_i(g_{ap}, m_a)$ function for each of the 20 simulations $i=1,...,20$. To determine the constraints, we find the contour curves that correspond to the 95$\%$ C.L., considering the appropriate degrees of freedom. The bounds for the 20 different realizations are shown in \cref{fig:ALPlimits} as red dotted curves, where the region inside each one is excluded. Then, we computed the pointwise mean, $\langle \chi^2(g_{ap}, m_a) \rangle$, and the pointwise standard deviation, $\sigma_{\chi^2}(g_{ap}, m_a)$, of the functions $\chi^2_i(g_{ap}, m_a)$. Thus, the combined constraint is determined by the contour curve that corresponds to the 95$\%$ C.L. of $\langle \chi^2(g_{ap}, m_a) \rangle$, while the 1$\sigma$ band is set by the 95$\%$ confident levels of $\langle \chi^2(g_{ap}, m_a) \rangle \pm \sigma_{\chi^2}(g_{ap}, m_a)$, shown in \cref{fig:ALPlimits} in purple.

As expected, there are significant variations among the constraints obtained from each individual simulation, which is reflected in the relatively large width of the resulting uncertainty band. This effect has a stronger impact on the bounds at ALP masses below $\sim 1$~MeV. For such low masses, the ALP spectrum peaks at $E_a\sim10$~MeV so most ALPs are emitted with ultra-relativistic velocities. This implies that the ALP flux diminishes, since only a small number of SNe contribute: their positions and explosion times must be tuned to obtain a significant ALP flux at Earth exactly during the observation window. This is a similar argument to the one used in Ref.~\cite{Alonso-Gonzalez:2024ems}, which limits the application of the smooth diffuse approximation to $m_a > 1$ MeV (shown in gray). In contrast, the numerical simulation developed here can be applied to any mass, which allows us to extend the constraint down to $m_a\sim 0.5$ MeV. Our method allows us to convincingly state that, below this mass, the flux has decreased so much that it is not enough to generate a measurable signal and the regions with $m_a\lesssim 0.5$ MeV are allowed.

\section{Fermionic DM}
\label{sec:darkphotons}

In Ref.~\cite{DeRocco:2019jti}, constraints on a DM model mediated by a heavy dark photon were derived from the lack of observation of a diffuse galactic DM flux in direct detection experiments. Here we revisit that derivation using the stochastic approach.

We consider a model where DM is a Dirac fermion, $\chi$, that interacts with the SM content through the four-fermion operator
\begin{equation}
    \mathcal{L}_\textrm{int}=\frac{e\epsilon g_d}{\Lambda^2}\bar\chi\gamma_\mu\chi J^\mu_\textrm{em},
\end{equation}
where $e$ is the electron charge, $J^\mu_\text{em}$ is the SM electromagnetic current and $\Lambda$ is a high-energy scale. This kind of interaction can be generated if DM is charged under a dark gauge boson (dark photon) of mass $m_{A'}=\Lambda$ with gauge coupling, $g_d$, which kinetically mixes with the SM via a mixing parameter $\epsilon$.

As in the ALP case, high temperatures and densities in the proto-NS provide the perfect environment to produce large amounts of sub-GeV dark fermions. To be consistent with the literature, we used the analytic SN profiles provided in the Appendix A of Ref.~\cite{DeRocco:2019jti}. The two main channels that generate dark fermions are electron-positron annihilation, $e^-e^+\rightarrow\bar\chi\chi$, and neutron-proton bremsstrahlung, $np\rightarrow np\bar\chi\chi$. In the parameter space relevant for our study, these fermions  couple strongly enough to the SM to become trapped diffusively in the vicinity of the proto-NS formed after core-collapse. This trapping is mainly driven by their scattering on free protons, which is inefficient at transferring energy to the dark fermions due to the large mass difference between the DM particles and the nucleons. However, dark fermions can also scatter with electrons and positrons, which enables thermal exchange with the SM plasma. Then, the dark fermion energy spectrum produced by a single SN can be approximated by a Fermi-Dirac distribution with temperature, $T_{\rm th}$, equal to the SN temperature at the radius of the thermal decoupling temperature, $r_{\rm th}$. Hence, the spectrum can be expressed as
\begin{equation}
    \frac{d N^2_\chi}{d E_{\chi} d t } = \frac{dN_\chi(m_{\chi}, y)}{dt} \left( \frac{E_{\chi}^{2}-m_{\chi}^{2}}{\exp(E_{\chi}/T_{\rm th})+1} \right) \left( \int_{m_\chi}^{\infty} \frac{E'^{2}-m_{\chi}^{2}}{\exp(E'/T_{\rm th})+1}\, dE' \right)^{-1} \, ,
    \label{eq:DFspectrum}
\end{equation}
where ${dN_\chi}/{dt}$ is the total flux of dark fermions produced by a single SN, whose values we take from the Monte Carlo simulation computed in Ref.~\cite{DeRocco:2019jti}. This flux depends on the dark fermion mass, $m_{\chi}$, and on the effective coupling parameter, $y$, that can be written as
\begin{equation}
    y = \epsilon^2 \alpha_D \left( \frac{m_{\chi}}{m_{A'}} \right)^4\ ,
\end{equation}
in terms of the fine-structure constant of the dark U(1) sector $\alpha_D = g_d^2/4\pi$.

\begin{figure}[!t]
  \centering
  \includegraphics[width=0.49\textwidth]{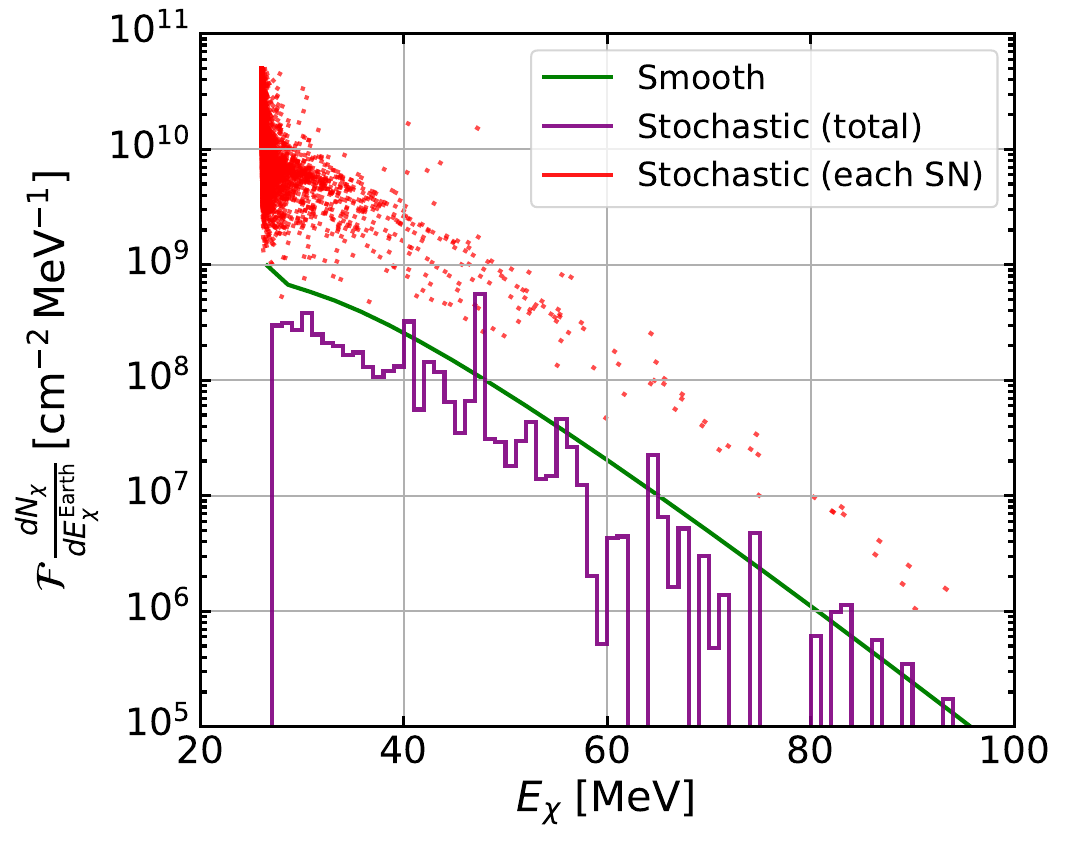}
  \raisebox{-1.0mm}{\includegraphics[width=0.49\textwidth]{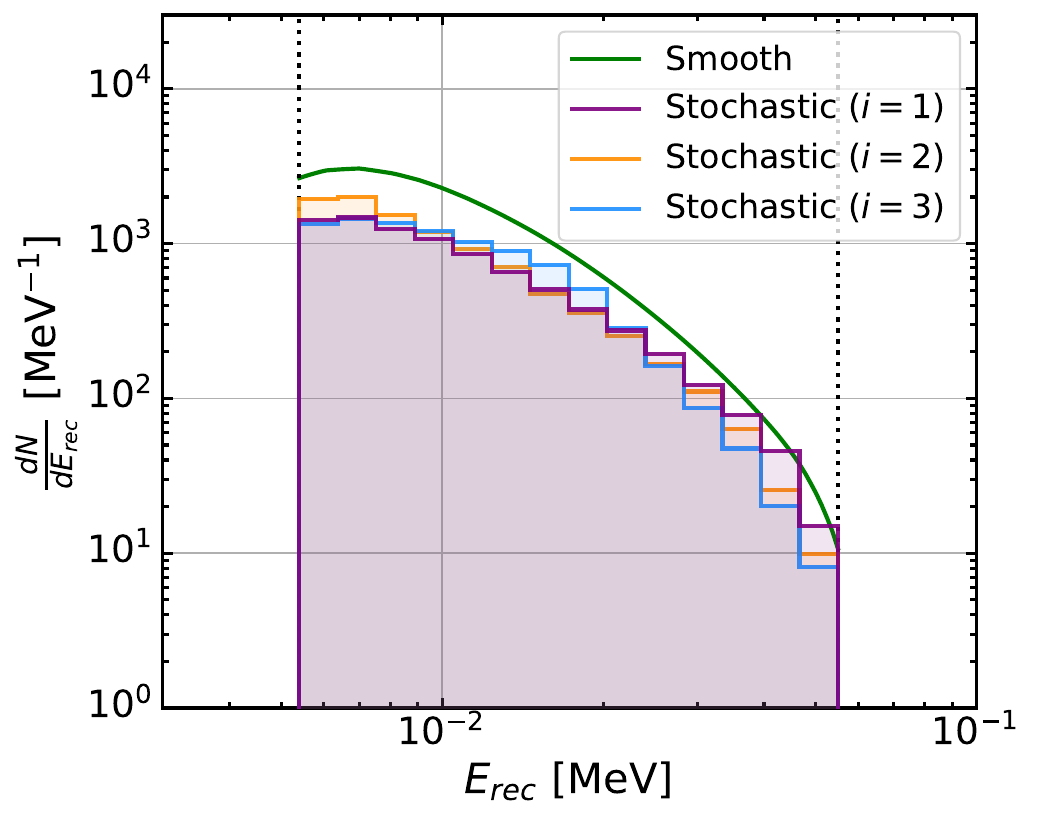}}
  \caption{Left: total fluence of $\chi$ at Earth computed with both the smooth and the stochastic approach for a single galactic SN history. Right: expected event rate in LZ with  a final exposure of 15~tonne-years, considering the smooth diffuse approximation (green line) and the stochastic approach (purple, orange and blue lines) for 3 different galactic SN history simulations. Both panels consider $m_{\chi}=26$ MeV, $\log(y)=-15.3$ and an observational time window of $\Delta t_{\rm obs}=2.74$~yr.
  The vertical dotted lines delimit the LZ energy region.} 
  \label{fig:fermionflux}
\end{figure}

Using the dark fermion spectrum produced by a single SN, we can determine the total particle fluence at Earth from past galactic SNe by inserting it into \cref{eq:diff_factor} and \cref{eq:stoch_factor2} for the smooth and stochastic approaches, respectively. The left panel of \cref{fig:fermionflux}, shows an example for $m_{\chi} = 26$~MeV, $\log(y) = -15.3$, and an observation time window of $\Delta t_{\rm obs} = 2.74$~yr. Since we consider the detection of the dark fermion flux in direct detection experiments, the time window corresponds to the 1000 days needed to reach an exposure of 15~tonne-years with LZ~\cite{LZ:2018qzl}. The short time window, compared to the SRP packet spread, confines the contribution of each SN to a very narrow, almost point-like, energy range (red curves). Nonetheless, the total stochastic fluence follows the same trend as the smooth diffuse approach.

The differential detection rate of the flux in liquid xenon detectors can be computed as
\begin{equation}
    \frac{d\, N}{dE_{\rm rec}}=N_t \int_{E_{\chi}^{\rm min}}dE_{\chi}^{\mathrm{Earth}}\frac{d\Phi_{\chi}}{dE_{\chi}^{\mathrm{Earth}}} \frac{d\, \sigma_{\chi {\rm Xe}}}{d\, E_{\rm rec}},
\end{equation}
where $E_{\chi}^{\mathrm{Earth}}=\alpha(r_{\rm th}) E_{\chi}$ is the observed dark fermion energy at Earth and $E_{\chi}$ its energy at production, i.e. at radius $r_{\rm th}$, $E_{\chi}^{\rm min}$ is the minimum dark fermion energy needed to produce a recoil of energy $E_{\rm rec}$. The differential flux ${d\Phi_{\chi}}/{dE_{\chi}^{\mathrm{Earth}}}$ is obtained by integrating \cref{eq:DFspectrum} over a 1.5 second interval assuming that the profiles remain unchanged during this period of time, and ${d\, \sigma_{\chi {\rm Xe}}}/{d\, E_{\rm rec}}$ is the differential dark fermion-xenon nucleus cross section (see Appendix B of Ref.~\cite{DeRocco:2019jti}).

In the right panel of \cref{fig:fermionflux}, we show the expected event rate at LZ (15~tonne-year exposure) considering $m_{\chi}=30$ MeV, $\log(y) = -15.3$, and 3 different galactic SN history simulations to illustrate the variability of the signal (the event rate for the 20 galactic SN histories used in this work can be found in Appendix~\ref{sec:extraplots}). We consider a similar analysis as performed in the WIMP search of Ref.~\cite{LZ:2024zvo}. Then, we compute the expected number of events as 
\begin{equation}
     N=\int_{E_{\rm rec}^{\rm min}}^{E_{\rm rec}^{\rm max}} \, dE_{\rm rec} \, \frac{d\, N}{dE_{\rm rec}} \, ,
\end{equation}
where the recoil energy range taken is $E_{\rm rec}=[5.4,55]$ keV.

\begin{figure}[!t]
  \centering
  \includegraphics[width=0.49\textwidth]{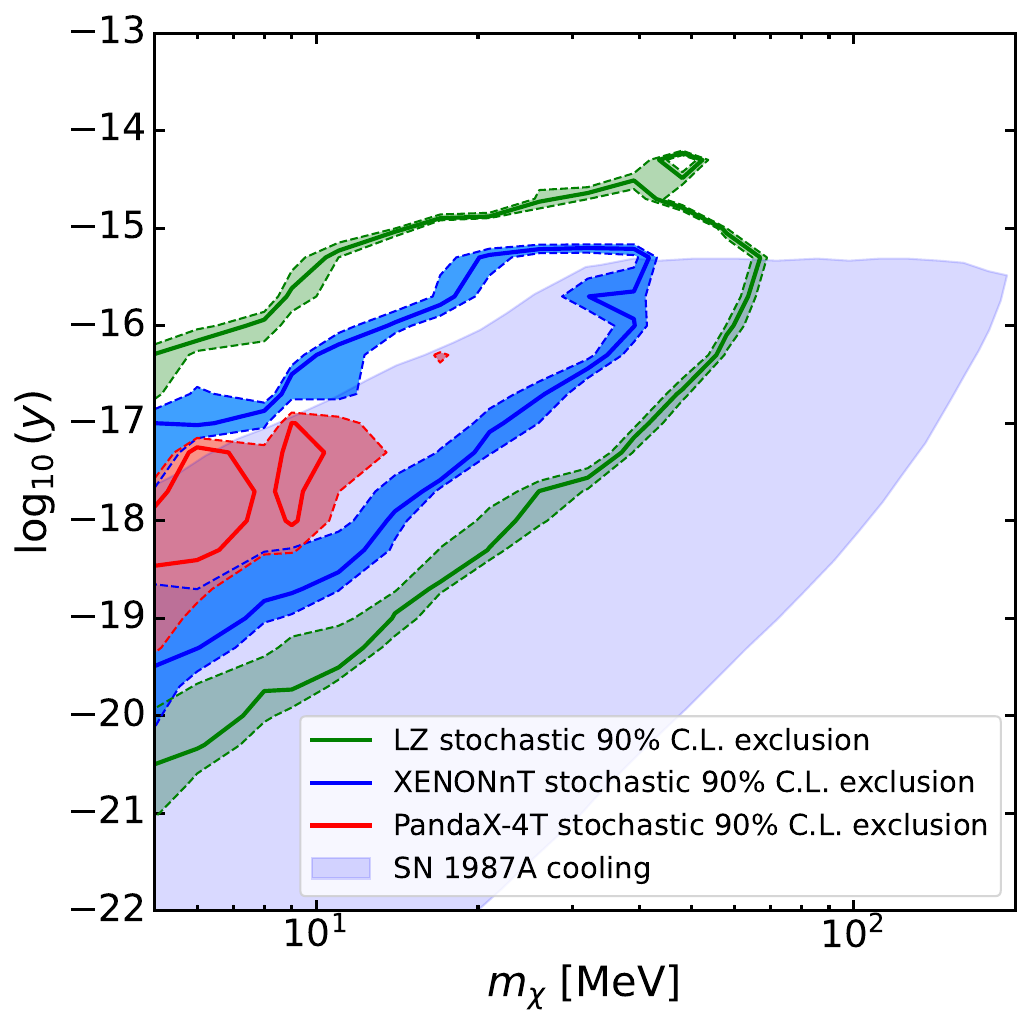}
  \includegraphics[width=0.49\textwidth]{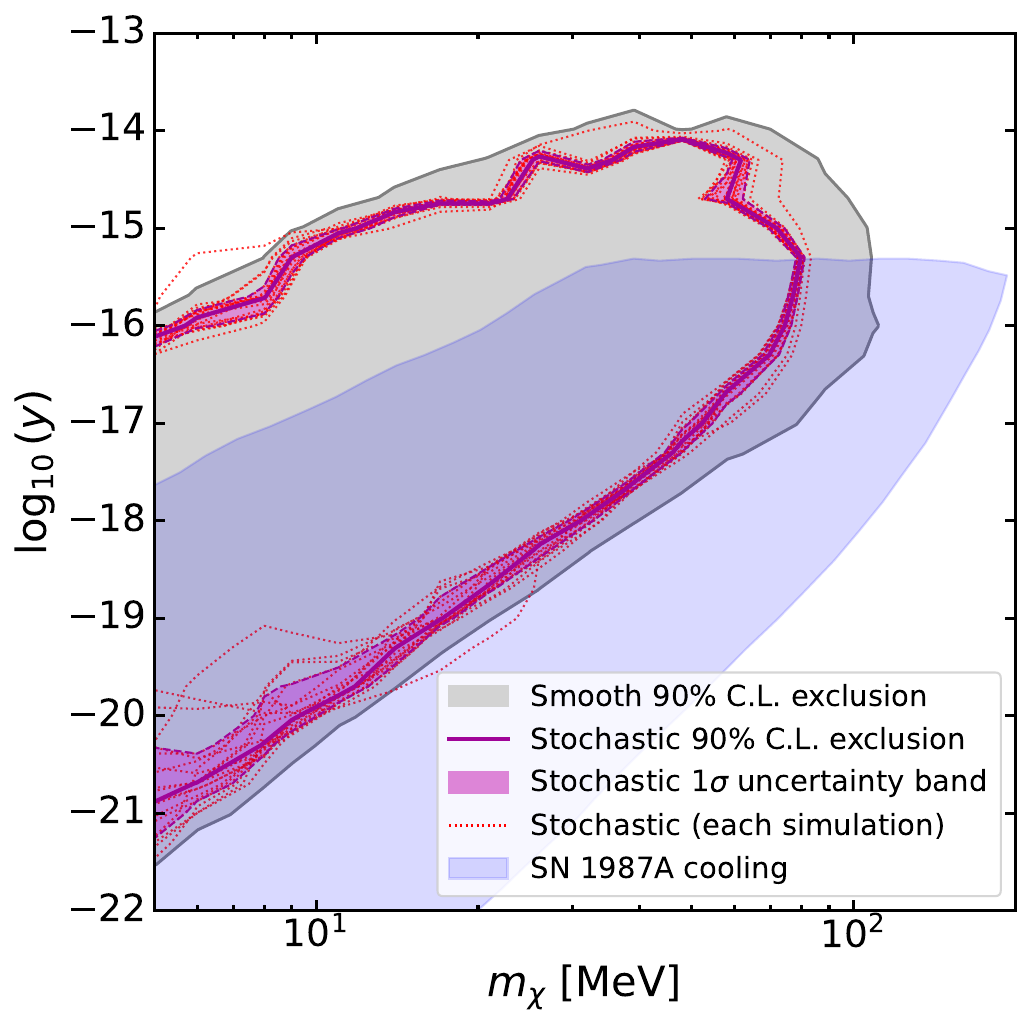}
  \caption{Left: Constraints on the $(y,\,m_\chi)$ parameter space from LZ (green) 4.2~tonne-year exposure~\cite{LZ:2024zvo}, XENONnT (blue) 3.1~tonne-year exposure~\cite{XENON:2025vwd}, and PandaX-4T (red) 1.54~tonne-year exposure~\cite{PandaX:2024qfu}, using the stochastic approach, with their corresponding 1$\sigma$ uncertainty bands.  Right: Projected bounds for LZ with 15~tonne-year exposure~\cite{LZ:2018qzl} on the fermionic DM parameter space, in gray using the smooth diffuse approximation, and in purple the stochastic approach, including a 1$\sigma$ uncertainty band due to the variations considering 20 SN galactic history simulations. In both panels, the light blue shaded area corresponds to the region excluded by cooling bounds from SN 1987A \cite{DeRocco:2019jti}.
  }
  
  \label{fig:fermionlimits}
\end{figure}

The left panel of \cref{fig:fermionlimits} shows the constraints on the dark fermion $(y, m_{\chi})$ parameter space derived from the experimental data of LZ 4.2 tonne-year exposure~\cite{LZ:2024zvo}, XENONnT 3.1 tonne-year exposure~\cite{XENON:2025vwd}, and PandaX-4T 1.54 tonne-year exposure~\cite{PandaX:2024qfu}. Each solid curve employs the pointwise mean expected number of events $\langle N \rangle$, with its $1\sigma$ band (dashed curves for each colour) computed as the pointwise standard deviation of $N_i$. We consider a Poisson likelihood-ratio test, then for each point in the parameter space we compute the test statistic
\begin{equation}
    q = 2 \left[N + B - N_{\rm obs} + N_{\rm obs} \, \text{ln}\left(\frac{N_{\rm obs}}{N+B}\right)\right] \, ,
\end{equation}
where $B$ denotes the expected background, and $N_{\rm obs}$ the observed number of events, both reported by each experiment. To determine the 90$\%$ C.L. exclusion limit, we find $N$ such that $\sqrt{q} = 1.28$. \footnote{Due to the low background in the nuclear recoil region of interest, the resulting contours are very similar to the background-free hypothesis, $N=2.3$.} The limit at $m_{\chi} = 5$ MeV is set by BBN constraints. Current searches using lower energy thresholds, LZ 5.7~tonne-year exposure~\cite{LZ:2025igz}, XENONnT 7.8~tonne-year exposure~\cite{XENON:2026qow}, PandaX-4T 1.2~tonne-year exposure~\cite{PandaX:2025rrz}, do not lead to constraints in the relevant parameter space.

Finally, on the right panel of \cref{fig:fermionlimits} we show the projected constraints for LZ with 15~tonne-year exposure~\cite{LZ:2018qzl}. The 90$\%$ C.L. exclusion is determined assuming a background-free scenario, which corresponds to finding the curve that satisfies $N=2.3$. In gray we show the bound obtained with the smooth approximation. The red dotted curves were computed with each SNe galactic history simulation, i.e., using $N_i$ with $i=1,...,20$. We can see that the parameter space that LZ would be able to explore and exclude if no significant excess is found is relaxed for the stochastic approach, mainly due to the loss of low-energetic ALPs, which fail to give a signal above threshold. As in the ALP case, the significant variations between simulations are reflected in the large uncertainty band, especially for low masses, $m_{\chi} \lesssim 10$ MeV.

\section{Conclusions}
\label{sec:conclusions}

In this article, we have revisited the assumption that the flux of semi-relativistic particles (SRPs) produced in galactic core-collapse supernovae (SNe) can be treated as a smooth diffuse flux at terrestrial experiments. This was based on the fact that these particles would travel to the Earth at different speeds, depending on the energies at which they are produced, resulting in the superposition of multiple packets with a typical time spread of centuries or even millennia for each SN. Here, we have critically examined this hypothesis, which ignores two important details.

First, due to the small observational window in typical experiments, only a narrow energy range in the spectrum for each SRP packet is observed (which depends on the SN distance and the SRP mass). Therefore, although many packets overlap at Earth, the resulting flux is not smooth. Second, for some combinations of SN distance, explosion time, and SRP mass, the SRPs of a given packet that would produce signals in the relevant energy range for the detector have already crossed the Earth. This results in a reduction of the total flux, an effect that is more important for light SRPs.

In order to properly incorporate these two features, in this article we have carried out a numerical simulation of the galactic history of core-collapse SNe over the past $9\times 10^5$~yr, assuming a distribution based on the galactic SN rate profile. For each SN, we have computed the flux of SRPs at Earth and its individual contribution to the total energy spectrum. By sampling the times and spatial locations of individual explosions, we have obtained the aggregated flux at Earth. We have observed that the resulting SRP flux exhibits sizeable fluctuations in the spectral shape, reflecting the stochastic nature of the SN population. This framework also allows us to consistently explore the low-mass regime, where the reduced temporal overlap among SN contributions limits the applicability of the smooth flux approximation.

While the smooth flux provides a reasonable estimate for deriving bounds on SRPs with masses above $1$~MeV, it does not accurately describe the signal expected at the detector. It systematically overestimates the expected flux by including contributions from SRP packets whose relevant energy range has already passed the Earth, leading to overly stringent bounds. The actual SRP signal is intrinsically non-smooth and extremely sensitive to the history of past SNe. As an application of this framework, we have revisited existing bounds in the literature for MeV axion-like particles and fermionic dark matter. We obtain constraints that are slightly weaker than previously reported.

This method provides a robust framework for computing SRP fluxes that can be applied to a wide range of models and experimental searches.

\section*{Acknowledgements}

We would like to thank Malcolm Fairbairn, Alessandro Lella, and Giuseppe Lucente for helpful discussions. ADP is supported by a Simons Foundation’s fellowship through the Targeted Grant to Instituto Balseiro. We acknowledge support from the Spanish Agencia Estatal de Investigaci\'on through the grants PID2024-155874NB-C22 and CEX2020-001007-S, funded by MCIN/AEI/10.13039/501100011033.

\appendix

\section{Signal variability}
\label{sec:extraplots}

\begin{figure}[!t]
  \centering
  \includegraphics[width=0.49\textwidth]{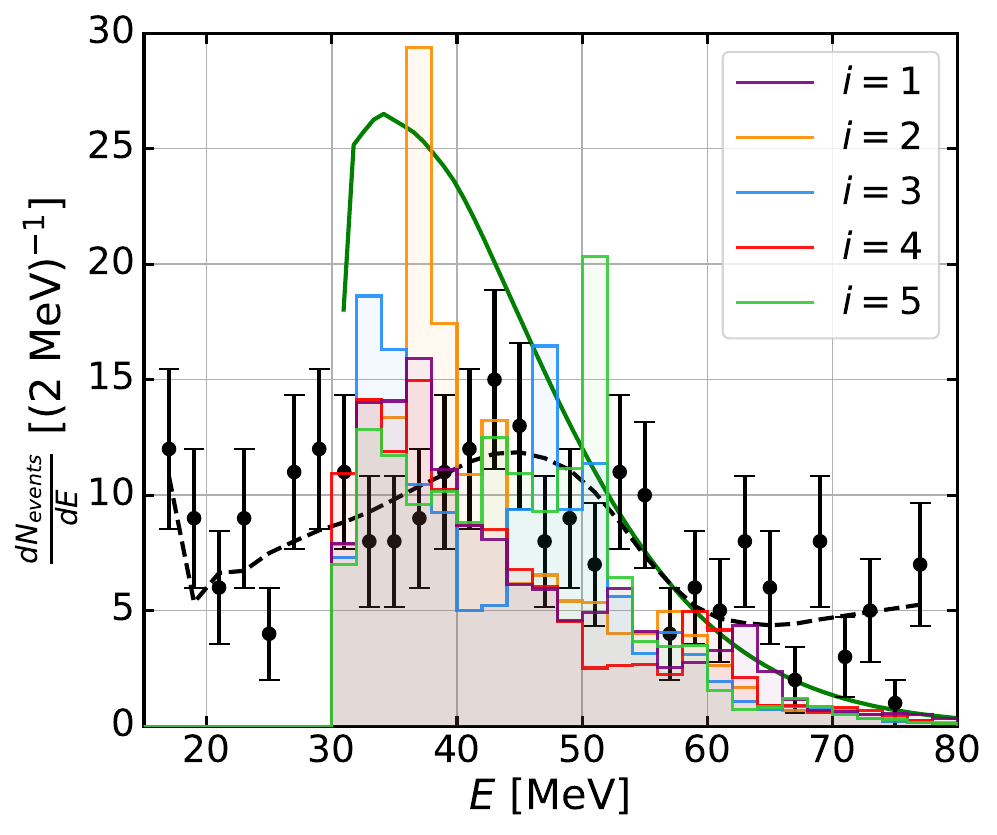}
  \includegraphics[width=0.49\textwidth]{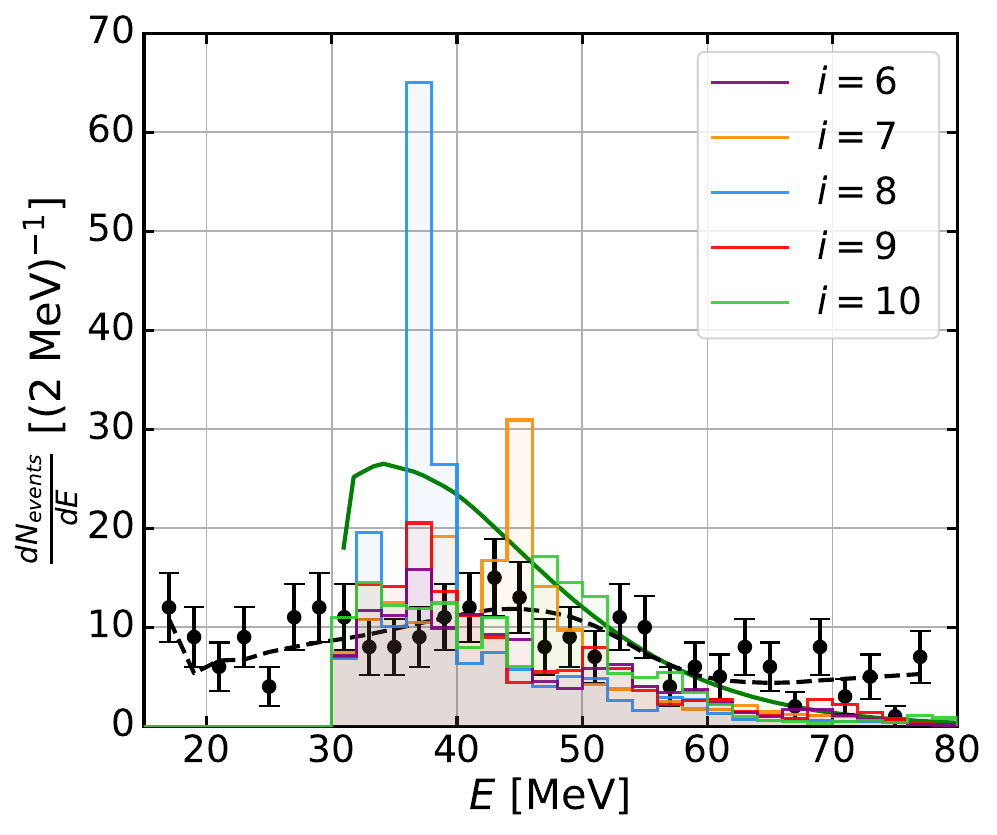}
  \includegraphics[width=0.49\textwidth]{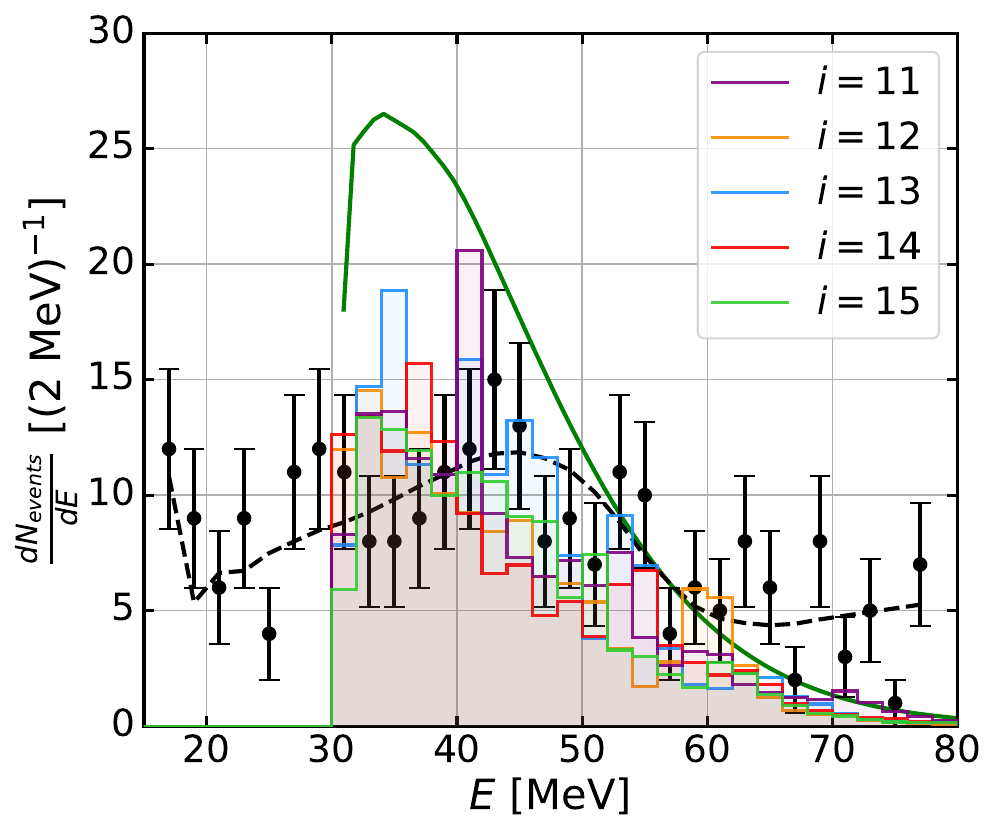}
  \includegraphics[width=0.49\textwidth]{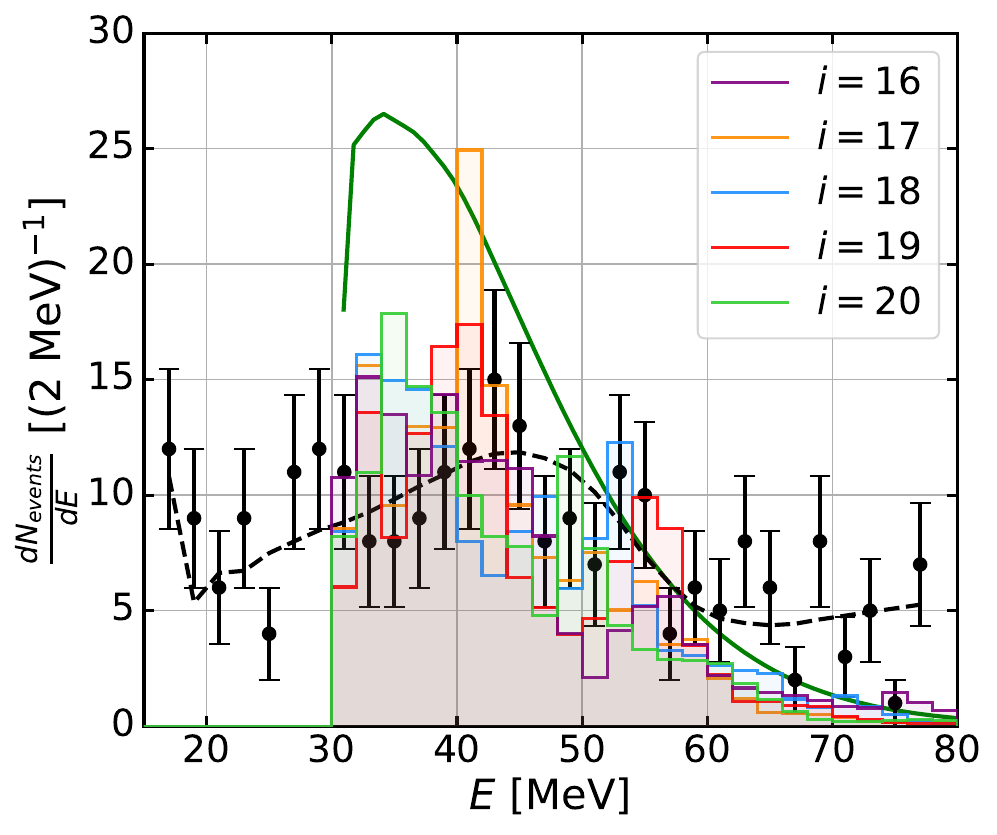}
  \caption{Expected event rate due to ALPs in SK phase IV for $i=1,...,20$ galactic SN history simulations as coloured histograms. The smooth diffuse event rate is shown as a dark green line, the expected events from SM processes in black dashed line and the observed events as black points with error bars, taken from  Ref.~\cite{Super-Kamiokande:2021jaq}.
  }
  \label{fig:ALP_extraplots}
\end{figure}

\begin{figure}[!t]
  \centering
  \includegraphics[width=0.49\textwidth]{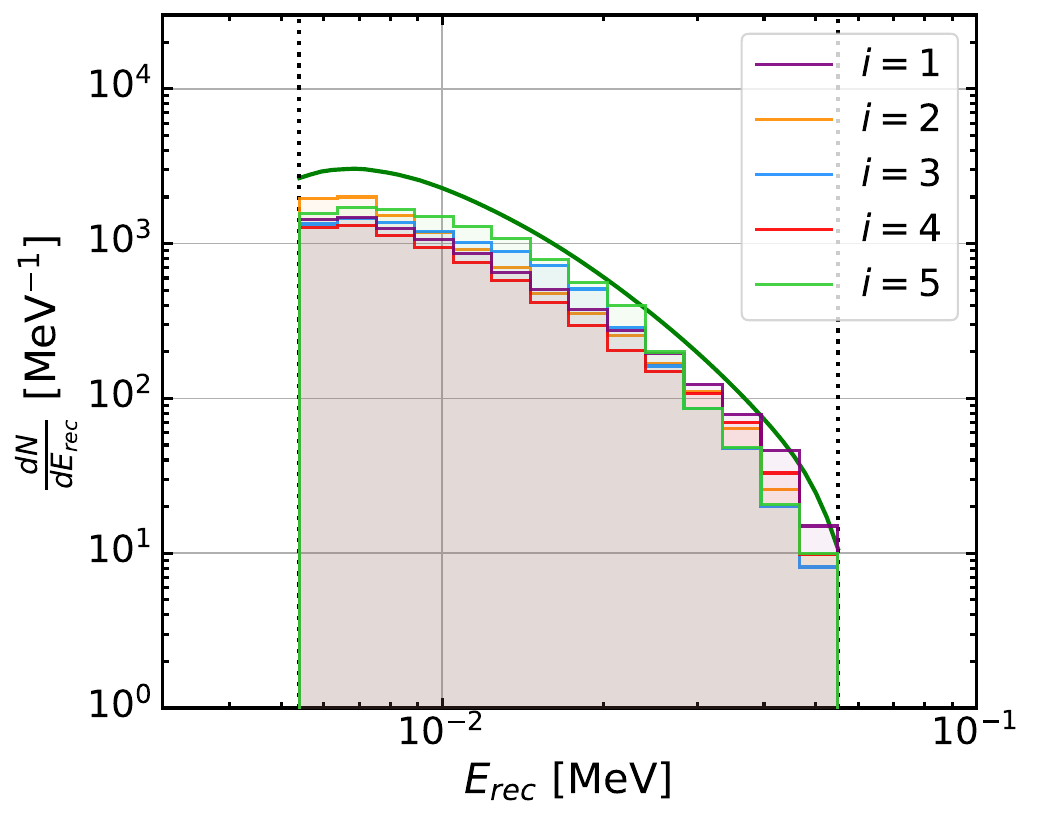}
  \includegraphics[width=0.49\textwidth]{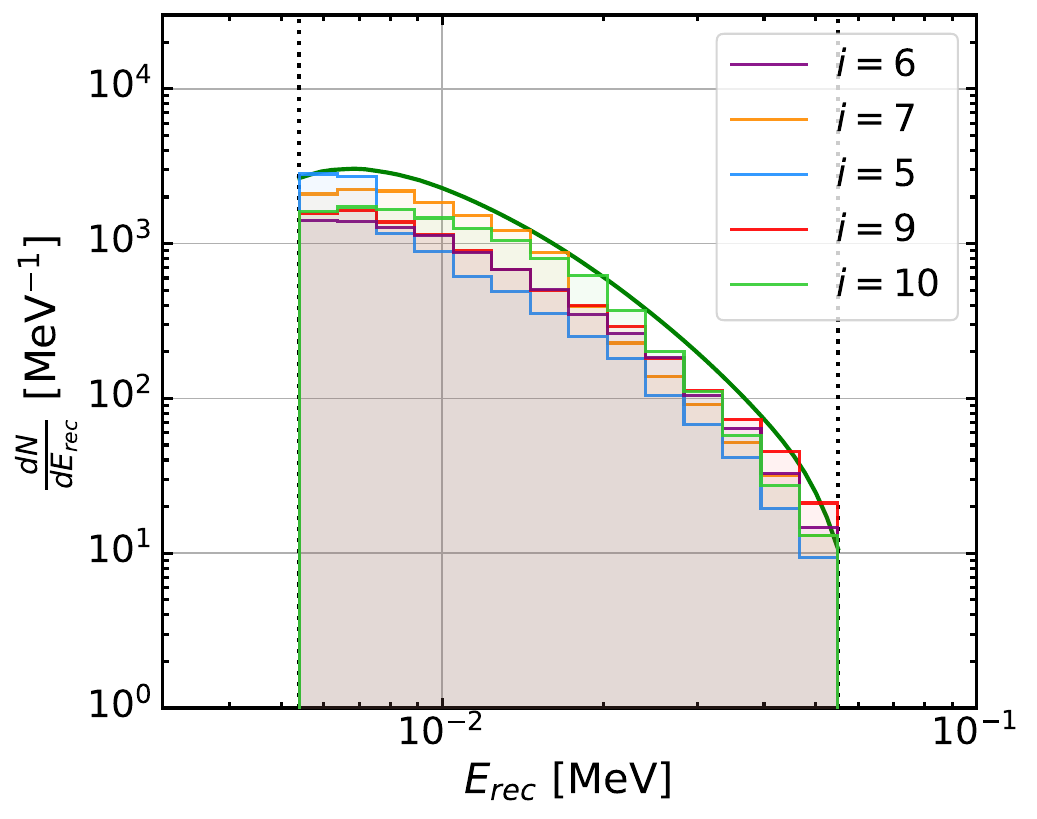}
  \includegraphics[width=0.49\textwidth]{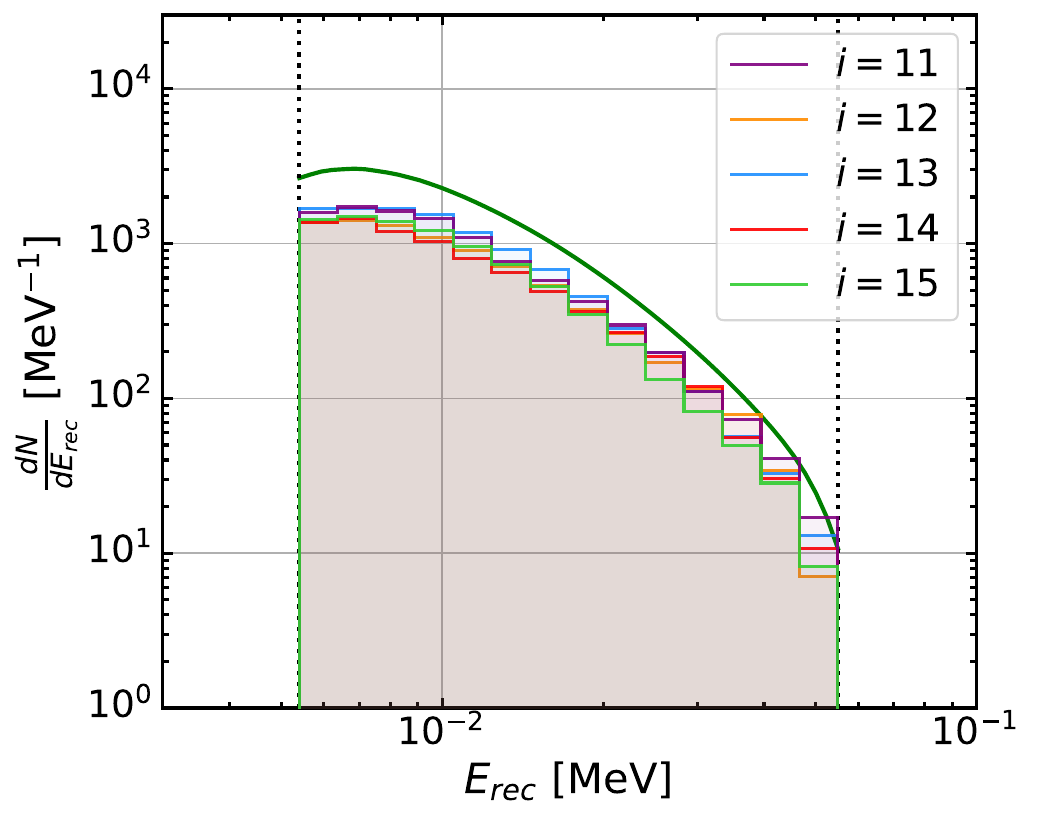}
  \includegraphics[width=0.49\textwidth]{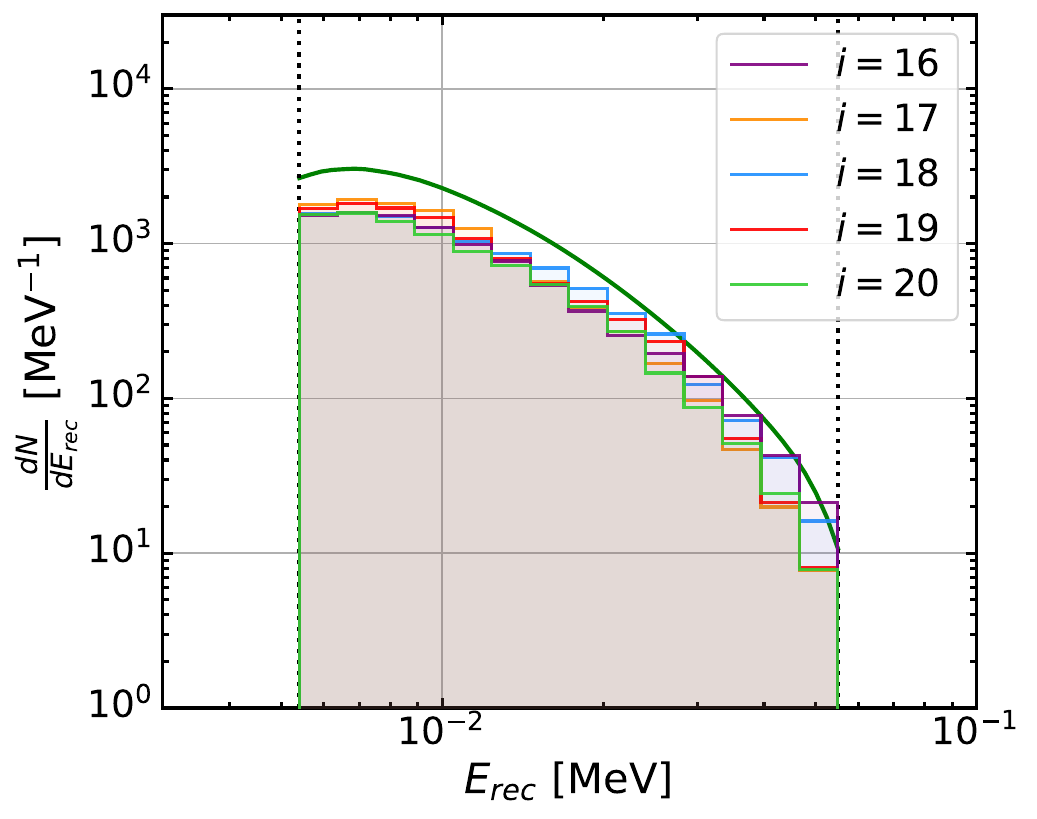}
  \caption{Expected event rate due to fermionic DM $\chi$ in LZ for $i=1,...,20$ galactic SN history simulations as coloured histograms. The smooth diffuse event rate is shown as a dark green line.
  }
  \label{fig:fermion_extraplots}
\end{figure}

In this appendix, we present the expected signal rate for the 20 simulations, individually. In Fig.~\ref{fig:ALP_extraplots} we show the ALP expected rate in SK phase IV and in Fig.~\ref{fig:fermion_extraplots} the fermionic DM case in LZ. For both models, significant differences appear between simulations, meaning that the signal strongly depends on the galactic SN  history.

\clearpage

\bibliographystyle{JHEP}
\bibliography{SGaSNoF.bbl}

\end{document}